\documentclass[12pt]{article}
\usepackage{graphicx}

\def \bea{\begin{eqnarray}}
\def \beq{\begin{equation}}

\def \eea{\end{eqnarray}}
\def \eeq{\end{equation}}

\def \ok{\overline{K}^0}

\def \s{\sqrt{2}}
\def \st{\sqrt{3}}

\def \thet{\theta_\eta}

\def \c{\circ}
\begin{document}
\rightline{EFI 10-18}
\rightline{arXiv:1008.4083}
\rightline{August 2010}
\bigskip
\centerline{\bf RELATIVE PHASES IN DALITZ PLOT AMPLITUDES}
\centerline{\bf FOR $D^0 \to K_S \pi^+ \pi^-$ AND $D^0 \to \pi^0 K^+ K^-$}
\bigskip

\centerline{Bhubanjyoti Bhattacharya\footnote{bhujyo@uchicago.edu} and
Jonathan L. Rosner\footnote{rosner@hep.uchicago.edu}}
\bigskip
\centerline{\it Enrico Fermi Institute and Department of Physics}
\centerline{\it University of Chicago, 5640 S. Ellis Avenue, Chicago, IL 60637}
\bigskip

\begin{quote}
Relative phases of amplitudes for $D$ meson decays to a light pseudoscalar
meson $P$ and a light vector meson $V$ decaying to two pseudoscalar mesons
will lead to characteristic interferences on the three-body Dalitz plot.
These phases may be compared with predictions of a flavor-symmetric treatment
which extracts contributing amplitudes and their relative phases from a fit
to $D \to PV$ decay rates.  Good agreement was found previously for the cases
of $B^0 \to K^+ \pi^- \pi^0$ and $D^0 \to \pi^+ \pi^- \pi^0$.  The present
work is devoted to the decays $D^0 \to K_S \pi^+ \pi^-$ and $D^0 \to \pi^0
K^+ K^-$, for which agreement is not found.  Several suggestions are offered
for this discrepancy.
\end{quote}

\leftline{PACS categories: 11.30.Hv, 13.25.Ft, 14.40.Lb}

\section{INTRODUCTION}

The relative phases of amplitudes for decays of charmed mesons to three
light pseudoscalar mesons $P$ are useful in extracting the phase
$\gamma$ of the Cabibbo-Kobayashi-Maskawa (CKM) matrix.  (See, e.g.,
\cite{Briere:2009aa} and references therein.)  These phases may be specified
by amplitude fits to Dalitz plots.  For amplitudes dominated by quasi-two-body
final states such as $PV$, where $V$ denotes a light vector meson, such phases
are also specified in fits to decay rates based on flavor symmetry.  (For
recent examples see Refs.\ \cite{Bhattacharya:2008ke,Cheng:2010ry,%
Bhattacharya:2010uy}.)

Good agreement between the two methods of extracting relative phases of
$D \to PV$ amplitudes was found previously for $B^0 \to K^+ \pi^- \pi^0$
\cite{Gronau:2010kq} and $D^0 \to \pi^+ \pi^- \pi^0$
\cite{Bhattacharya:2010id}.  In the present work we investigate a similar
question for the decays $D^0 \to K_S \pi^+ \pi^-$ \cite{Briere:2009aa,%
Aubert:2005iz,Aubert:2008bd,Poluektov:2010wz,delAmoSanchez:2010xz} and
$D^0 \to \pi^0 K^+ K^-$ \cite{Cawlfield:2006hm,Aubert:2007dc}. For these two
processes, we do not find agreement between phases based on Dalitz plot
analyses and those calculated from our flavor-symmetric amplitude analysis.
(The decays $D^0 \to K_S K^+ K^-$ are also studied in many of these references,
but the important role of scalar resonances, for which quark-model assignments
are uncertain, puts a similar analysis beyond our reach for the moment.)

We recall notation for amplitudes in the flavor-symmetric analysis and quote
their values obtained from previous fits \cite{Bhattacharya:2008ke,
Bhattacharya:2010uy} in Sec.\ II.  We then construct the amplitudes for
relevant $D \to PV$ subprocesses in Sec.\ III, and compare them with those
extracted from Dalitz plot fits in Sec.\ IV.  We discuss possible reasons for
the observed discrepancies in Sec.\ V, and conclude in Sec.\ VI.  An Appendix
discusses some phase conventions.

\section{AMPLITUDES FROM PREVIOUS FITS}

We recall the notation from our previous analyses of $D \to PV$ decays
\cite{Bhattacharya:2008ke}.  The ratios of singly-Cabibbo-suppressed (SCS)
amplitudes to Cabibbo-favored (CF) ones, and of doubly-Cabibbo-suppressed
(DCS) to SCS ones, are SCS/CF = DCS/SCS = $\tan \theta_C \equiv \lambda =
0.2305$ \cite{BM}, with $\theta_C$ the Cabibbo angle and signs governed by the
relevant CKM factors.

\begin{figure}
\mbox{\includegraphics[width=0.46\textwidth]{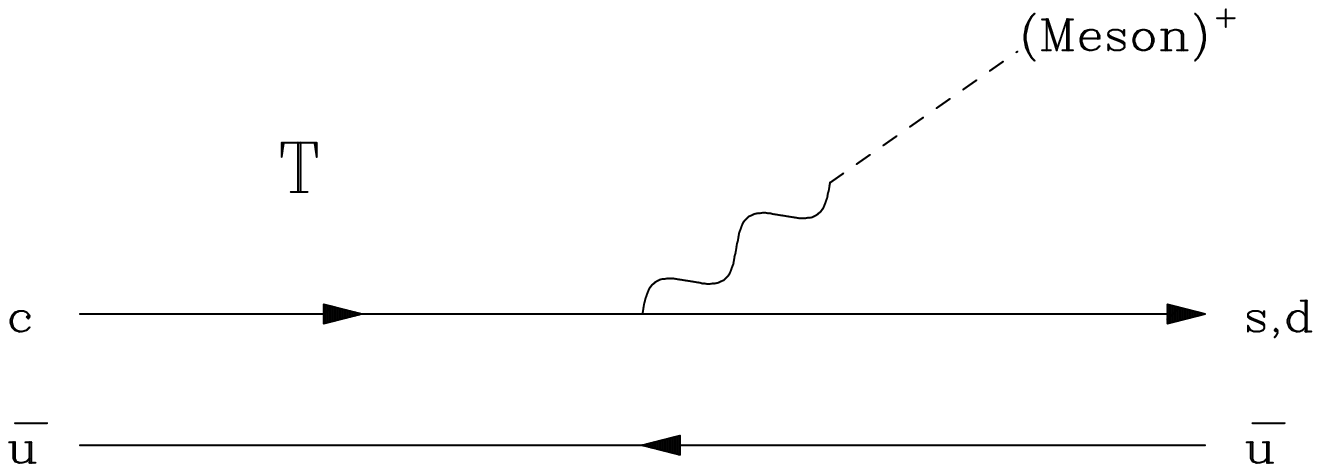} \hskip 0.3in
      \includegraphics[width=0.46\textwidth]{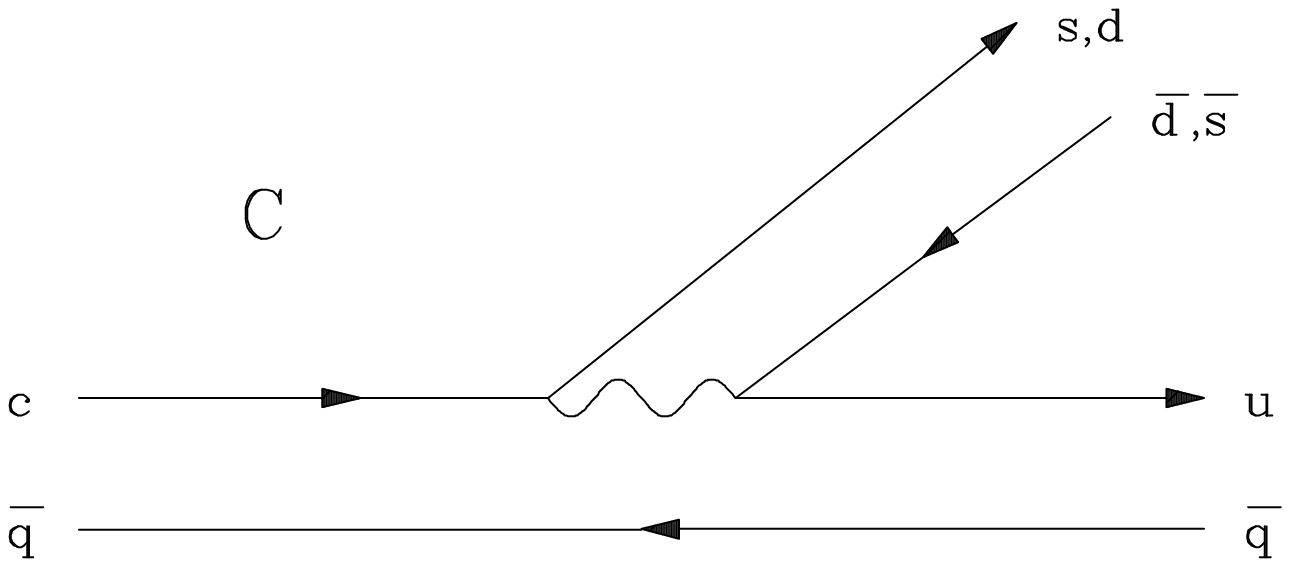}}
\vskip 0.3in
\mbox{\includegraphics[width=0.46\textwidth]{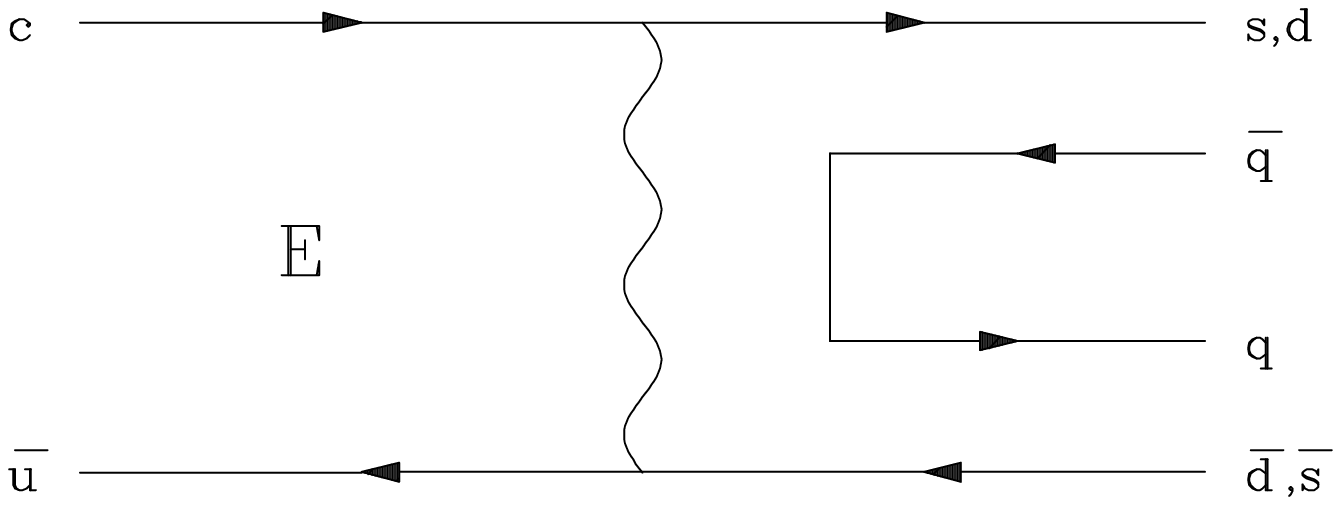} \hskip 0.3in
      \includegraphics[width=0.46\textwidth]{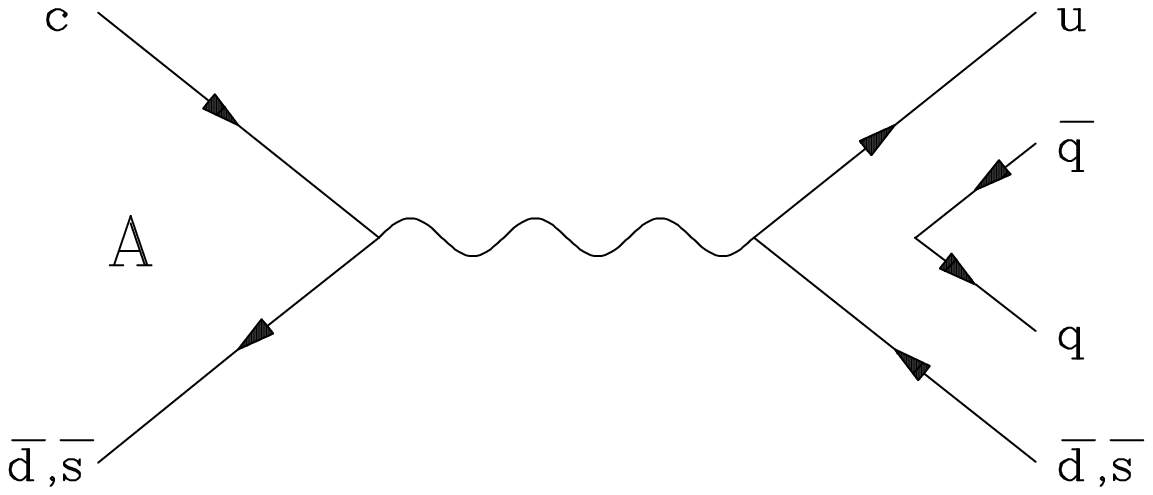}}
\caption{Flavor topologies for describing charm decays.  $T$: color-favored
tree; $C$: color-suppressed tree; $E$ exchange; $A$: annihilation.
\label{fig:TCEA}}
\end{figure}

For present purposes we shall be interested in amplitudes labeled as $T$
(``tree''), $C$ (``color-suppressed''), and $E$ (``exchange''), illustrated in
Fig.\ \ref{fig:TCEA}.
For $PV$ final
states, a subscript on the amplitude denotes the meson ($P$ or $V$) containing
the spectator quark.  The partial width $\Gamma(H \to PV)$ for the decay of a
heavy meson $H$ is expressed in terms of an invariant amplitude ${\cal A}$ as
\beq
\Gamma(H \to PV) = \frac{p^{*3}}{8 \pi M_H^2}|{\cal A}|^2~,
\eeq
where $p^*$ is the center-of-mass (c.m.) 3-momentum of each final particle,
and $M_H$ is the mass of the decaying particle.  With this definition, the
amplitudes ${\cal A}$ are dimensionless.

Fits to rates for $D \to PV$ Cabibbo-favored decays not involving $\eta$ or
$\eta'$ provide information on the amplitudes $T_V$, $C_P$, and $E_P$, as
shown in Table \ref{tab:tvcpep}.  For the amplitudes $T_P$, $C_V$, and $E_V$,
one needs information on the $\eta$-$\eta'$ mixing, and Table \ref{tab:tpcvev}
shows results for two values $\thet = 19.5^\circ$ and $11.7^\circ$.

\section{CONSTRUCTION OF $D \to PV$ AMPLITUDES}

The $D^0 \to PV$ amplitudes of interest for the present discussion are shown in
Tables \ref{tab:amps19} ($\thet = 19.5^\circ$) and \ref{tab:amps11} ($\thet =
11.7^\circ$), along with their representations in the flavor-SU(3) language and
their values.

\begin{table}[t]
\caption{Solution in Cabibbo-favored charmed meson decays to $PV$ final states
favored by fits \cite{Bhattacharya:2008ke} to singly-Cabibbo-favored decays.
\label{tab:tvcpep}}
\begin{center}
\begin{tabular}{c c c} \hline \hline
   $PV$   &  Magnitude  &   Relative \\
amplitude & ($10^{-6}$) & strong phase \\ \hline
$T_V$ & $3.95 \pm 0.07$ & --- \\
$C_P$ & $4.88 \pm 0.15$ & $\delta_{C_PT_V} = (-162 \pm 1)^\circ$ \\
$E_P$ & $2.94 \pm 0.09$ & $\delta_{E_PT_V} = (-93  \pm 3)^\circ$ \\
\hline\hline
\end{tabular}
\end{center}
\end{table}

\begin{table}[h]
\caption{Solutions for $T_P$, $C_V$, and $E_V$ amplitudes in Cabibbo-favored
charmed meson decays to $PV$ final states, for $\eta - \eta'$ mixing angle of
$\thet = 19.5^\c$ and $11.7^\circ$.
\label{tab:tpcvev}}
\begin{center}
\begin{tabular}{c c c c c} \hline \hline
 & \multicolumn{2}{c}{$\thet=19.5^\c$} & \multicolumn{2}{c}{$\thet=11.7^\c$} \\
$PV$ &  Magnitude  &  Relative  &  Magnitude  &  Relative \\
ampl.& ($10^{-6}$) & strong phase & ($10^{-6}$) & strong phase \\ \hline
$T_P$ & 7.46$\pm$0.21 & Assumed 0 & 7.69$\pm$0.21 & Assumed 0 \\
$C_V$ & 3.46$\pm$0.18 &$\delta_{C_VT_V} = (172 \pm 3)^\c$ & 4.05$\pm$0.27 &
 $\delta_{C_VT_V} = (162 \pm 4)^\c$ \\
$E_V$ & 2.37$\pm$0.19 &$\delta_{E_VT_V} =(-110 \pm 4)^\c$ & 1.11$\pm$0.22 &
 $\delta_{E_VT_V} =(-130 \pm 10)^\c$ \\ \hline\hline
\end{tabular}
\end{center}
\end{table}

Two notable relations in these tables are
\bea \label{rel1}
A(D^0 \to K^{*-} K^+) & = & \lambda A(D^0 \to K^{*-} \pi^+)~~,~~~\\
\label{rel2} A(D^0 \to K^{*+} K^-) & = & -(1/\lambda) A(D^0 \to K^{*+} \pi^-)~~,
\eea
independent of the fitted values of these parameters.  The phases of the
first two amplitudes are expected to be equal, while the second pair should
have a relative phase of $180^\circ$.  Taking the quotient of the two, one
finds
\beq \label{eqn:ratio}
\frac{A(D^0 \to K^{*-} K^+)}{A(D^0 \to K^{*+} K^-)} = - \lambda^2
\frac{A(D^0 \to K^{*-} \pi^+)}{A(D^0 \to K^{*+} \pi^-)} \label{rel3}
\eeq
This relation can be checked using relative phases of Dalitz plot amplitudes,
and will be one of the tests performed in Section.\ V.

\section{COMPARISON WITH MAGNITUDES AND\\ PHASES IN DALITZ PLOT ANALYSES}

The amplitudes for the $D^0 \to PV$ processes described in the previous Section
do not have any information about the vector meson decay. In order to obtain
information about how these amplitudes interfere on the $D^0 \to K_S \pi^+
\pi^-$ or $D^0 \to \pi^0 K^+ K^-$ Dalitz plots, we need to let the vector
meson decay to two-pseudoscalar final states. In a $D^0 \to ABC$ Dalitz plot,
if we consider the intermediate process $D^0 \to RC$ where $R$ is the
intermediate $AB$ resonance, we need to multiply the amplitude for the
intermediate process by the appropriate isospin Clebsch--Gordan factor
governing the decay of the vector meson.  The spin part of this amplitude on
the Dalitz plot is $T = -2 \vec p_A \cdot \vec p_C$ where $\vec p_i$
is the 3-momentum of the particle $i$ in the resonance rest frame.  This
implies that if we switch the order of the particles $A$ and $B$ in the vector
meson decay, then the phase of the resulting amplitude changes by $\pi$.
Thus the order of the two pseudoscalar mesons in the vector meson decay is
crucial for our understanding of interferences on the Dalitz plot. Here we are
guided by a set of conventions kindly communicated to us by R. Andreassen
\cite{RApc} and K. Mishra \cite{Mishra}. These conventions are noted in Table
\ref{tab:conv}.  The particle index numbers in column 5 of this table show the
order of the two pseudoscalars from the vector meson decay, for each process
with an intermediate vector resonance. The respective Clebsch--Gordan factors
indicate the weight attached to the $D^0 \to PV$ amplitude to obtain its
contribution to the interferences on the corresponding Dalitz plot.

\begin{table}
\caption{Amplitudes for $D^0 \to PV$ decays of interest for the present
discussion (in units of $10^{-6}$).  The first three processes contribute
to $D^0 \to K_S \pi^+ \pi^-$, while the last three contribute to $D^0 \to \pi^0
K^+ K^-$.  Here we have taken $\thet=19.5^\circ$.
\label{tab:amps19}}
\begin{center}
\begin{tabular}{c c c c c c} \hline \hline
$D^0$ final & Amplitude & \multicolumn{4}{c}{Amplitude $A$} \\
  state  & representation & Re & Im & $|A|$ & Phase ($^\circ$) \\ \hline
$K^{*-} \pi^+$ & $T_V+E_P$               &   3.796 & --2.936 & 4.799 & -- 37.7 \\
$\rho^0 \ok$ & $(C_V-E_V)/\s$       & --1.850 &   1.915 & 2.663 &   134.0 \\
$K^{*+} \pi^-$ & $-\lambda^2(T_P + E_V)$ & --0.353 &   0.118 & 0.373 &   161.5 \\
$K^{*-} K^+$ & $\lambda(T_V+E_P)$        &   0.875 & --0.677 & 1.106 & -- 37.7 \\
$K^{*+} K^-$ & $\lambda(T_P+E_V)$        &   1.533 & --0.513 & 1.616 & -- 18.5 \\
$\phi \pi^0$ & $\lambda C_P/\s$          & --0.756 & --0.246 & 0.795 & --162.0 \\
\hline \hline
\end{tabular}
\end{center}
\end{table}

\begin{table}
\caption{Same as Table \ref{tab:amps19} but for $\thet=11.7^\circ$.
\label{tab:amps11}}
\begin{center}
\begin{tabular}{c c c c c c} \hline \hline
$D^0$ final & Amplitude & \multicolumn{4}{c}{Amplitude $A$} \\
  state  & representation & Re & Im & $|A|$ & Phase ($^\circ$) \\ \hline
$K^{*-} \pi^+$ & $T_V+E_P$               &   3.796 & --2.936 & 4.799 & -- 37.7 \\
$\rho^0 \ok$ & $(C_V-E_V)/\s$            & --2.219 &   1.486 & 2.671 &   146.2 \\
$K^{*+} \pi^-$ & $-\lambda^2(T_P + E_V)$ & --0.371 &   0.045 & 0.373 &   173.1 \\
$K^{*-} K^+$ & $\lambda(T_V+E_P)$        &   0.875 & --0.677 & 1.106 & -- 37.7 \\
$K^{*+} K^-$ & $\lambda(T_P+E_V)$        &   1.608 & --0.196 & 1.620 & --  6.9 \\
$\phi \pi^0$ & $\lambda C_P/\s$          & --0.756 & --0.246 & 0.795 & --162.0 \\
\hline \hline
\end{tabular}
\end{center}
\end{table}

\begin{table}
\caption{Conventions for order of the two pseudoscalars in vector meson decay
\cite{RApc,Mishra}. Here we choose the CP-even state $K_S = \frac{K^0 - \ok}
{\s}$ following the convention in Ref.\ \cite{Bhattacharya:2008ss}.
\label{tab:conv}}
\begin{center}
\begin{tabular}{c c c c c c} \hline \hline
Dalitz Plot & \multicolumn{2}{c}{Bachelor Particle} & \multicolumn{3}{c}{Vector Meson Decay} \\
       & Meson & Index & Process & Indices & Clebsch factor \\ \hline
       & $K_S$   & 1 & $\rho^0 \to \pi^+ \pi^-$ & 23 & 1          \\
$D^0 \to K_S \pi^+ \pi^-$& $\pi^+$ & 2 & $K^{*-} \to K_S \pi^-$ & 13
 & $\sqrt{2/3}$   \\
   & $\pi^-$ & 3 & $K^{*+} \to K_S \pi^+$   & 12 & $-\sqrt{2/3}$ \\ \hline
   & $\pi^0$ & 1 & $\phi   \to K^+ K^-$     & 23 & $1/\s$     \\
$D^0 \to \pi^0 K^+ K^-$  & $K^+$   & 2 & $K^{*-} \to K^- \pi^0$ & 31
 & --$1/\st$    \\
   & $K^-$   & 3 & $K^{*+} \to \pi^0 K^+$   & 12 & $- 1/\st$  \\ \hline \hline
\end{tabular}
\end{center}
\end{table}

We fix the amplitudes of $D^0 \to \rho^0 K_S$ for $D^0 \to K_S \pi^+ \pi^-$
and $D^0 \to K^{*+} K^-$ for $D^0 \to \pi^0 K^+ K^-$ to 1.0 and obtain the
amplitudes and phases of the other processes relative to these. The results
are shown for $\thet = 19.5^\circ$ in Table \ref{tab:Kpipi} for $D^0 \to K_S
\pi^+ \pi^-$ and in Table \ref{tab:piKK} for $D^0 \to \pi^0 K^+ K^-$. One
does not see agreement between the relative phases predicted in the flavor
-SU(3) approach and those implied by the Dalitz plot analysis. A similar set
of results may be obtained using $\thet = 11.7^\circ$. Even though there are
slight changes in the relevant amplitudes and phases, they still do not agree
with those obtained from the Dalitz plot analysis. In the next Section we
shall discuss some possible reasons for this discrepancy.

\begin{table}
\caption{Relative amplitudes and phases in $D^0 \to K_S \pi^+ \pi^-$ Dalitz
plot. We use $\thet = 19.5^\c$.
\label{tab:Kpipi}}
\begin{center}
\begin{tabular}{c c c c c} \hline \hline
 Decay & \multicolumn{2}{c}{Relative Theoretical}
       & \multicolumn{2}{c}{Relative Experimental}
  \cite{delAmoSanchez:2010xz}   \\
 mode & Amplitude & Phase ($^\circ$) & Amplitude & Phase ($^\circ$) \\ \hline
$D^0 \to \rho^0 K_S$   & 1.0      & 0.0      & 1.0        & 0.0     \\
$D^0 \to K^{*-} \pi^+$ & 1.472$\pm$0.187 & 188$\pm$7 & 1.735$\pm$0.005
 & 133.5$\pm$0.2 \\
$D^0 \to K^{*+} \pi^-$ & 0.114$\pm$0.015 & 27$\pm$7 & 0.164$\pm$0.003
 & --44.0$\pm$1.1 \\

\hline \hline
\end{tabular}
\end{center}
\end{table}

\begin{table}
\caption{Relative amplitudes and phases in $D^0 \to \pi^0 K^+ K^-$ Dalitz plot. We use
$\thet = 19.5^\c$.
\label{tab:piKK}}
\begin{center}
\begin{tabular}{c c c c c}
\hline \hline
 Decay & \multicolumn{2}{c}{Relative Theoretical}
       & \multicolumn{2}{c}{Relative Experimental}
  \cite{Aubert:2007dc}   \\
  mode & Amplitude & Phase ($^\circ$) & Amplitude & Phase ($^\circ$) \\ \hline \hline
$D^0 \to K^{*+} K^-$ & 1.0      & 0.0     & 1.0          & 0.0       \\
$D^0 \to K^{*-} K^+$ & 0.685$\pm$0.049 & --19$\pm$3 & 0.601$\pm$0.016
 &--37 $\pm$2.9 \\
$D^0 \to \phi \pi^0$ & 0.602$\pm$0.042 & 37$\pm$3& 0.690$\pm$0.022
 &--20.7$\pm$16.5 \\ \hline \hline
\end{tabular}
\end{center}
\end{table}

\section{POSSIBLE SOURCES OF DISCREPANCIES}

\subsection{Inaccuracies or instabilities of the flavor-SU(3) approach}

Although the flavor-SU(3) approach has had some success in fitting rates
of charm decays to $PP$ and $PV$ \cite{Bhattacharya:2008ke,Cheng:2010ry,%
Bhattacharya:2010uy,Bhattacharya:2008ss,Bhattacharya:2009ps}, as well as
in describing relative phases in $D^0 \to K^- \pi^+ \pi^0$ \cite{Gronau:2010kq}
and $D^0 \to \pi^+ \pi^- \pi^0$ \cite{Bhattacharya:2010id} Dalitz plots,
there are some notable shortcomings.  Perhaps the most familiar is the
prediction of equal rates for $D^0 \to K^+ K^-$ and $D^0 \to \pi^+ \pi^-$,
whereas the former rate is about 2.8 times the latter \cite{Amsler:2008}.
One possibility is that an earlier fit to Cabibbo-favored decay rates
\cite{Rosner:1999,Chiang:2002mr} with $|C_P| < |T_V|$ (``Solution B'') in
Ref.\ \cite{Bhattacharya:2008ke}, rejected because it did not fit SCS rates
as well as the amplitudes quoted in Table \ref{tab:tvcpep}, nevertheless has
some validity. One may use the B1 solutions quoted in Ref.\
\cite{Bhattacharya:2008ke} and check the relative amplitudes and phases as was
done for the A1 solutions in the previous section.  It turns out that the B1
solutions don't agree with relative amplitude predictions in the case of $D^0
\to \pi^0 K^+ K^-$, indicating that $|C_P| < |T_V|$ is not very helpful.  In
the case of $D^0 \to K_S \pi^+ \pi^-$, where the amplitudes do not explicitly
depend on $C_P$, the B1 solutions give reasonable results for relative
amplitudes, but fail to agree with the relative phases.

Even though a flavor-SU(3) approach might predict equal strong phases for a
pair of amplitudes, it has been pointed out that this relation could be
violated by SU(3)-breaking effects \cite{Wolf:1995}.  Thus, although
flavor SU(3) predicts equal strong phases for $D^0 \to K^- \pi^+$ and
$D^0 \to K^+ \pi^-$, there might be no reason to expect such an equality.
In fact, this is an experimental question, which can be attacked by a
variety of means \cite{Gronau:2001nr}.  The CLEO Collaboration has addressed
this problem using tagged $D^0$ mesons produced in pairs at the $\psi(3770)$
resonance, and finds a strong phase consistent with zero \cite{TQCA}.

\subsection{Dalitz plot conventions}

It is notoriously tricky to specify conventions for Dalitz plot amplitudes for
$D \to RC \to ABC$ when considering vector mesons $R=V$, because of the
importance of choosing the order correctly in $V \to AB$ decay.  This
question was found to be non-trivial in both cases mentioned above in which
agreement between flavor SU(3) and Dalitz plot analyses was eventually found.
One cross-check which should be relatively airtight is the comparison of
relative phases implied by Eqs.\ (\ref{rel1}) and (\ref{rel2}). Here we use a
modified version of Eq.\ (\ref{rel3}):
\beq
\frac{A(D^0 \to K^{*-} K^+)}{A(D^0 \to K^{*+} K^-)} = -\,r\,e^{\it{i}\,\phi}\,
\lambda^2 \frac{A(D^0 \to K^{*-} \pi^+)}{A(D^0 \to K^{*+} \pi^-)},
\eeq
where $r$ and $\phi$ determine the amplitude and phase of the deviation from
Eq.\ (\ref{rel3}). [Eq.\ (\ref{rel3}) corresponds to $r = 1$ and $\phi =
0^\circ$.] We find the following results from the experimental values, taking
into account the signs and magnitudes of the Clebsch-Gordan factors in
Table \ref{tab:conv} and remembering the present convention $K_S = (K^0 -
\ok)/\s$:
\beq
r_{\rm ex} = 1.069 \pm 0.034~;~~\phi_{\rm ex} = (-34.5 \pm 3.1)^\circ~.
\eeq
The experimental result differs from $0^\circ$ by a phase $\Delta
\phi_{\rm ex} = -34.5^\circ$, which is small enough that any remaining
discrepancy is probably at least not due to a sign convention.

Comparison of the experimental and theoretical values for the two sides of
Eq.\ (\ref{eqn:ratio}) can indicate whether a sign might be misplaced
in the conventions for $D^0 \to \pi^0 K^+ K^-$ (left-hand side) or $D^0 \to K_S
\pi^+ \pi^-$ (right-hand side).  In Table \ref{tab:ratiocomp} we compare the
ratios in theory (based on Table \ref{tab:amps19}) and experiment (based
on Tables \ref{tab:conv}, \ref{tab:Kpipi}, and \ref{tab:piKK}).

\begin{table}
\caption{Comparison between predicted and measured ratios in Eq.\
(\ref{eqn:ratio}).
\label{tab:ratiocomp}}
\begin{center}
\begin{tabular}{c c c c c} \hline \hline
Amplitude & \multicolumn{2}{c}{Predicted} & \multicolumn{2}{c}{Measured} \\
 ratio    & Magnitude & Phase ($^\circ$) & Magnitude & Phase ($^\circ$) \\
\hline
$A(K^{*-}K^+)/A(K^{*+}K^-)$ & 0.685$\pm$0.049 &--19.2$\pm$2.2 & 0.601$\pm$0.016
& --37.0$\pm$2.9 \\
$-\lambda^2 A(K^{*-}\pi^+)/A(K^{*+}\pi^-)$ & 0.685$\pm$0.049 & --19.2$\pm$2.2
& 0.562$\pm$0.010 & --2.5$\pm$1.1 \\ \hline \hline
\end{tabular}
\end{center}
\end{table}

We have checked the phase conventions for $D^0 \to \pi^0 K^+ K^-$ using the
fact that in Ref.\ \cite{Cawlfield:2006hm} the $\bar K^*$ and $K^*$ resonances
are found to interfere destructively with one another.  In that reference and
in the BaBar analysis \cite{Aubert:2008bd}, this would follow from the
conventions stated in Table \ref{tab:conv}, which have been confirmed to be
those used \cite{Mishra}.  These involve cyclic permutations of the particle
indices for the three $VP$ cases.  On the other hand, approximate agreement
of our relative phase prediction for $D^0 \to K^{*-} \pi^+$ and $D^0 \to
K^{*+} \pi^-$ with that measured experimentally \cite{delAmoSanchez:2010xz}
suggested to us that the conventions for $D^0 \to K_S \pi^+ \pi^-$ were as
stated in Table \ref{tab:conv}, and did {\it not} involve cyclic particle
indices.  This was also confirmed to be the case \cite{RApc}. (See Appendix.)

\subsection{Existence of multiple solutions to Dalitz plot fits}

It is possible to find multiple solutions of relative phases on Dalitz plots.
Usually one chooses the ``best-fit'' solution, but such a choice may depend
on assumptions about other amplitudes in the fit.  To that end, we suggest
that fits be attempted starting from the relative phases we have proposed
in Table \ref{tab:Kpipi} and \ref{tab:piKK}.

\subsection{Alternative parametrization of S-wave $K \pi$ amplitudes}

The relative phases of different $D^0 \to PV$ amplitudes in $D^0 \to K_S
\pi^+ \pi^-$ and $D^0 \to \pi^0 K^+ K^-$ Dalitz plots depend on the
interference of these amplitudes with those involving $K \pi$ in an
S-wave final state.  This is particularly important for $D^0 \to K_S \pi^+
\pi^-$, as the $PV$ bands in the Dalitz plot do not overlap with one another.

The BaBar analyses we have quoted (e.g., \cite{Aubert:2008bd,%
delAmoSanchez:2010xz,Aubert:2007dc}) parametrize $K \pi$ S-wave amplitudes
using a form consistent with LASS data \cite{LASS}.  The elastic scattering
amplitudes in these data are expected to rise linearly with squared
center-of-mass energy $s$ as a result of the Adler zero
\cite{Adler:1965,Weinberg:1966} at $s \simeq
m_K^2 - m_\pi^2/2$ \cite{Bugg:2003,Bugg:2006,Bugg:2009}.  Because of this
behavior, a pole for a scalar resonance (``$\kappa$'') will appear at a much
lower mass than in a na\"{\i}ve Breit-Wigner parametrization
\cite{DescotesGenon:2006uk}.  An inelastic process such as $D \to \bar K \pi
\pi$ does not involve the Adler zero, so a description of the S-wave $K \pi$
amplitude via the LASS amplitude is not appropriate.  A similar difference
between elastic $\pi \pi$ amplitudes consistent with current algebra,
unitarity, and crossing symmetry \cite{Brown:1968,Brown:1971} (which contain
the Adler zero) and inelastic processes such as $\gamma \gamma \to \pi^+ \pi^-$
\cite{Goble:1972, Goble:1989} (which do not) is responsible for the peaking at
very low dipion effective mass of the cross section for the latter process.

Although we cannot at this point indicate the quantitative effect a
different $K \pi$ S-wave amplitude parametrization would have on the
Dalitz plot analyses, it has been shown in fits to $D \to \bar K \pi \pi$
decays \cite{Bugg:2006} that one obtains an improved description of the
data by introducing the Adler zero into the $K \pi$ amplitude.  A similar
exercise would be worthwhile in the BaBar, Belle, and CLEO data, especially in
light of the importance of relative phases in three-body $D^0$ decays for
determining phases of the CKM matrix \cite{Briere:2009aa,Giri:2003ty}.

\section{CONCLUSIONS}

We have compared experimental determinations of relative phases in Dalitz
plot amplitudes for $D^0  \to K_S \pi^+ \pi^-$ and $D^0 \to \pi^0 K^+ K^-$
with predictions from a flavor-SU(3) approach.  In contrast to the previously-%
studied cases of $B^0 \to K^+ \pi^- \pi^0$ \cite{Gronau:2010kq} and
$D^0 \to \pi^+ \pi^- \pi^0$ \cite{Bhattacharya:2010id}, we do not find
agreement.  A simpler test, Eq.\ (\ref{rel3}) relating the ratio
$A(D^0 \to K^{*-} K^+)/A(D^0 \to K^{*+} K^-)$ to the ratio $A(D^0 \to K^{*-}
\pi^+)/A(D^0 \to K^{*+} \pi^-)$ with coefficient $- \lambda^2 = -0.053$, is
satisfied by magnitudes of amplitudes but fails in phase by
$(34.5 \pm 3.1)^\circ$.  The phase discrepancies in various relations seem
to be limited to less than $60^\circ$, suggesting that at least sign
conventions have been properly identified.  Three remaining possibilities for
these shortcomings include (a) inaccuracies of the flavor-SU(3) approach, (b)
the possibility that other Dalitz plot solutions exist with phases closer to
the flavor-SU(3) predictions, and (c) sensitivity to the parametrization of
the S-wave $K \pi$ amplitudes, in particular to the distinction between elastic
amplitudes which have an Adler zero and inelastic ones which do not.

\section*{ACKNOWLEDGMENTS}

We thank M. Dubrovin, M. Gaspero, K. Mishra, B. Meadows, and A. Soffer for
helpful
communications.  JLR is grateful to the Fermilab Theory Group for hospitality
during the completion of this study.  This work was supported in part by the
United States Department of Energy through Grant No.\ DE FG02 90ER40560.

\section*{APPENDIX:  PHASE CONVENTIONS IN THREE-BODY DECAYS}

\begin{figure}
\begin{center}
\includegraphics[width=0.6\textwidth]{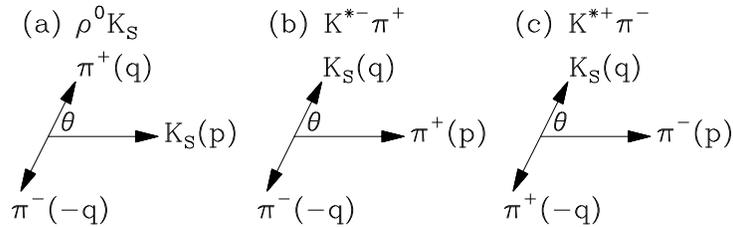}
\end{center}
\caption{Phase conventions of Ref.\ \cite{RApc} for the decays
$D^0 \to K_S \pi^+ \pi^-$.
\label{fig:Kpipi}}
\end{figure}

\begin{figure}
\begin{center}
\includegraphics[width=0.6\textwidth]{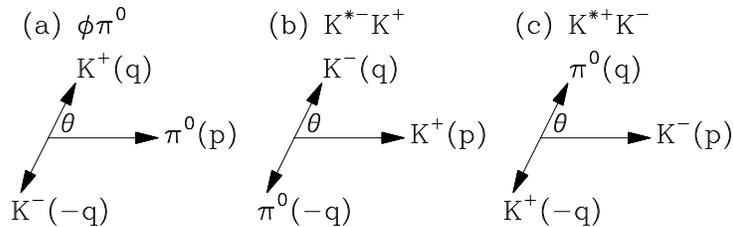}
\end{center}
\caption{Phase conventions of Ref.\ \cite{Mishra} for the decays
$D^0 \to \pi^0 K^+ K^-$.
\label{fig:piKK}}
\end{figure}

Here we expand upon the phase conventions defined in Table \ref{tab:conv}
\cite{RApc,Mishra}.  For a decay $D^0 \to RC \to (AB)C$, where $R$ is the
intermediate $AB$ resonance (here, a vector meson), the conventions for
$D^0 \to K_S \pi^+ \pi^-$ and $D^0 \to \pi^0 K^+ K^-$ are illustrated in
Figs.\ \ref{fig:Kpipi} and \ref{fig:piKK}, respectively.
The matrix element
$T = - 2 \vec{p}_A \cdot \vec{p}_C$ for the vector meson contribution to the
Dalitz plot amplitude for a decay $D \to ABC$ can also be expressed in terms
of masses and two-body effective masses
\beq
m^2_{AB} = (p_A+p_B)^2~,~~
m^2_{AC} = (p_A+p_C)^2~,~~
m^2_{BC} = (p_B+p_C)^2
\eeq
in the form (see, e.g., \cite{Kopp:2000gv,Bonvicini:2008jw}, containing also
expressions for a spin-2 resonance $R$):
\beq
- 2 \vec{p}_A \cdot \vec{p}_C = \frac{1}{2} \left( m^2_{AC} - m^2_{BC} \right)
+ \frac{1}{2 m^2_{AB}} \left( m_B^2 - m_A^2 \right) \left( m_D^2 - m_C^2
\right)~.
\eeq

The $K^*$ bands in $D^0 \to K_S \pi^+ \pi^-$ do not overlap with one another,
so one cannot directly see the interference between the Cabibbo-favored decay
$D^0 \to K^{*-} \pi^+$ and the doubly-Cabibbo-suppressed decay $D^0 \to
K^{*+} \pi^-$.  However, the $K^*$ bands do overlap in $D^0 \to \pi^0 K^+ K^-$,
and are of comparable strength, as both represent singly-Cabibbo-suppressed
decays.  It was noted in Ref.\ \cite{Rosner:2003} that the sign of the
interference can be readily diagnosed by inspecting the overlap region.
In fact, in Ref.\ \cite{Cawlfield:2006hm} the interference between the $K^{*+}$
and $K^{*-}$ bands was found to be destructive in the overlap region.  This
conclusion was based on enhancement of the $K^{*+}$ band in the low-$m(K^-
\pi^0)$ region, but suppression of the $K^{*-}$ band in the low-$m(K^+ \pi^0)$
region, indicating opposite signs of interference with a slowly-varying S-wave
component.

Referring to Fig.\ \ref{fig:piKK}, one sees that the low-$m(K^+\pi^0)$ region
in the $K^{*-}$ band [illustrated by the configuration (b)] and the
low-$m(K^- \pi^0)$ region in the $K^{*+}$ band [illustrated by the
configuration (c)] are defined with the same sign.  In that case, one
expects the relative phase between the $D^0 \to K^{*+} K^-$ and $D^0
\to K^{*-} K^+$ amplitudes on the $D^0 \to \pi^0 K^+ K^-$ Dalitz plot,
quoted in Table \ref{tab:piKK}, to be near $180^\circ$, as predicted
theoretically.  Taking account of the relative sign of the Clebsch-Gordan
coefficients $\pm 1/\st$ in the last column and last two rows of Table
\ref{tab:conv}, this means that one expects the phase of the ratio
on the left-hand side of Eq.\ (\ref{eqn:ratio}), $A(D^0 \to K^{*-} K^+)/
A(D^0 \to K^{*+} K^-)$, to be near zero, as predicted.  (See Table
\ref{tab:ratiocomp}.)

\end{document}